# Report on Mega-Constellations to the Government of Canada and the Canadian Space Agency

Independent Working Group of Astronomers

This document provides recommendations to the Government of Canada and the Canadian Space Agency (CSA) in response to their call for feedback on the future of Canadian space exploration. We focus on how the construction and long-term placement of mega-constellations of satellites into Earth orbit will affect astronomy and the view of the night sky by all peoples, with attention to all Canadians. We also highlight several environmental concerns associated with the construction and maintenance of these mega-constellations. The recommendations address ways that Canada can mitigate some of these negative effects.

This report is written by an independent working group that was commissioned by the Joint Committee on Space Astronomy of the Canadian Astronomical Society - Société Canadienne D'Astronomie (CASCA). The recommendations are based on views of and research conducted by astronomers within the Canadian astronomical community, but are not intended to reflect complete agreement among all astronomers within the community.

We have kept the recommendations general for reasons of flexibility. All recommendations are given with full commentary to provide examples of scope and, in some cases, potential ways of implementation. The recommendations are followed by a broader context that includes a sampling of the many issues that mega-constellations will cause and is not intended to be all-encompassing. These recommendations should be considered along with other recent planning and policy documents, including the Canadian Astronomy Long Range Plan 2020[1] and the IAU Dark and Quiet Skies Report[2].

In drafting the recommendations, we take the approach that space needs to be developed sustainably. In this regard, we use the Brundtland Report's definition[3]: "Sustainable development is development that meets the needs of the present without compromising the ability of future generations to meet their own needs". Thus, all recommendations here are made with the intent of minimizing the negative consequences of mega-constellations, while also recognizing that their development will continue. These recommendations are further written with the following in mind:

1. Mega-constellations have potential to aid society through, for example, enhanced connectivity of remote communities and search and rescue capabilities.
2. Canada is not a launching state.
3. Canada controls licensing for access to the Canadian market.
4. Canadian companies are participating in the development of mega-constellations.
5. Canada has a reputation as being an honest broker in international relations.
6. Canada plays a major role in the global astronomical research community.

---

[1] https://casca.ca/wp-content/uploads/2020/12/LRP2020_December2020-1.pdf
[2] https://www.iau.org/static/publications/dqskies-book-29-12-20.pdf
[3] Brundtland et al. (1987). At:
https://sustainabledevelopment.un.org/content/documents/5987our-common-future.pdf

# Recommendations

1. Canada should prioritize at national and international levels implementing the recommendations in the IAU Dark & Quiet Skies report, or versions iterated upon in a multilateral forum, to the extent that is feasible.

*Commentary: The Dark & Quiet Skies report[4] was written in collaboration between multiple organizations, including the International Astronomical Union (IAU) and the United Nations Office of Outer Space Affairs (UNOOSA). The report covers several concerns related to preserving the sky for humanity, including the negative effects of artificial light at night on human health and wildlife; radio interference; the impacts of low-Earth Orbit (LEO) satellites on the night sky; and loss of fundamental science opportunities. The report will be presented in April 2021 to the Science and Technology Subcommittee of the United Nations Committee on the Peaceful Uses of Outer Space (UN COPUOS). In particular, its Chapter 6 focuses on impacts of satellite mega-constellations on the night sky and on optical and radio astronomy. This Chapter includes 40 recommendations, divided into sections aimed at different interest groups. Ten of these recommendations are explicitly for national science funding agencies, policy makers and regulators, national standards agencies, national economic and space policy makers, and international policy makers. The corresponding recommendations are appended to this report for convenience.*

2. Canada should with all due speed update its Licencing of Space Stations (LSS) requirements to appropriately evaluate requests by mega-constellation satellite companies to broadcast in Canada, and to identify externalities.

*Commentary: Licencing requirements, including renewals, need to consider fully the impact of mega-constellations on the sky and the Earth and Space environments. This includes an operator's contribution to the total number of satellites visible at twilight and night, out-of-band radio emission, re-entry ground risks, re-entry atmospheric pollution, launch pollution, constellation-wide on-orbit fragmentation risks, and scientific loss at ground-based facilities. Environmental impact assessments that address, inter alia, the above issues and are appropriate for mega-constellations could be made a requirement for initial licences and renewals, as well as implementing interim reviews. Examples include (1) re-evaluating the first-come, first-serve policy[5] in light of de facto occupation of orbital shells and the wide-spread effects of radio spectrum use; (2) spectrum allocation and utilization compliance (cf. LSS3.2.3[4]), with the possibility of using interim report filings by operators to demonstrate that compliance; (3) the orbital debris mitigation guidelines, with attention to potential all-of-LEO effects and shorter de-orbiting times (cf. LSS3.3.3 and 25 yr rule); (4) the*

---

[4] https://www.iau.org/static/publications/dqskies-book-29-12-20.pdf
[5] CPC-2-6-02 -- Licensing of Space Stations. Available at:
https://www.ic.gc.ca/eic/site/smt-gst.nsf/eng/sf01385.html

> *Conditions of Licences (cf. LSS4.2); and, (5) the post-authorization reporting requirements (cf. LSS5). New guidelines for LSS might need to be developed, such as assessments for atmospheric pollution and potentially hazardous and non-hazardous re-entries of satellites, boosters, and their debris over Canadian territory and surrounding waters.*
>
> *Section 3.5.3 of the Dark & Quiet Skies report deals with the cultural significance of the night sky, which is a vital part of traditional knowledge for many peoples, including Indigenous Canadians[6]. Indigenous ways of knowing are based upon the land and sky context, and passing on that knowledge to future generations requires a clear view of the stars. The development of space affects a resource, the night sky, that is an important part of the cultures of many Indigenous Canadian groups. Recommendation 92.i. in the Truth and Reconciliation Commission Calls to Action is that corporations should engage in meaningful consultation and obtain free, prior, and informed consent of Indigenous peoples before proceeding with economic development projects. This recommendation should apply to corporations launching mega-constellations that affect the night sky over Canada, and consultations should be part of licensing requirements. It is true that many remote Indigenous communities and rural residents can benefit from access to the internet, but the launching corporations have not transparently given the general public enough information to understand the ramifications of mega-constellation construction.*

3. Canada should develop national policy and regulatory mechanisms to address externalities associated with mega-constellations.

> *Commentary: The negative effects of mega-constellations are shared by everyone on Earth, regardless of their use or ability to access services from them. Externalities include damage to viewing the night sky, atmospheric pollution and chemistry alteration, and risks to people and property from satellite, booster, and debris re-entries. Annual fees are already required as part of the licensing requirements (LSS4.5). Additional fees or fines under these regulations, or some other appropriately identified mechanism, such as a benefit sharing regime, might be considered for satellite operators who exceed specified metrics. Such metrics might include orbit and spectrum use thresholds, as well as the degree to which satellites exceed a specified visual brightness and the number that do so. The resulting fee structure or benefit sharing can in turn be used to aid in funding mitigation efforts for sky and environmental damages, as well as to help communities who are most affected by the loss of the night sky.*

---

[6] A few examples of Indigenous Canadian astronomy are contained in the following links:
https://science.ucalgary.ca/rothney-observatory/community/first-nations-skylore
https://www.ictinc.ca/blog/a-brief-look-at-indigenous-star-names https://www.nativeskywatchers.com/
http://www.virtualmuseum.ca/edu/ViewLoitLo.do?method=preview&lang=EN&id=5185
https://www.mcnallyrobinson.com/9781927849460/wilfred-buck/tipiskawi-kisik

IWGA Report on Mega-Constellations4. Canada should take a leadership role, through multilateral forums, in negotiating an international standard for addressing externalities associated with mega-constellations.

*Commentary: Canada is not a launching state, which limits the impact that national initiatives highlighted in Recommendation 3 might have on the sustainable development of space. Moreover, national initiatives cannot be so restrictive that benefits to Canadians from mega-constellations are limited by such policy (e.g., lack of service or excessive end-user costs), while the externalities remain. International standards must be developed and widely adopted. Canada could lead in this process given its success in previous efforts, such as the Montreal Protocol, and given its potential to be heavily affected by mega-constellation externalities compared with other uses.*

*As shown in Figure 3, the satellite orbital distribution will place some of the highest satellite concentrations over Canadian territory in terms of number of satellites per latitudinal band. These concentrations are a result of the inclination distribution of satellites. A concern is whether Canada will see a disproportionate externality burden as a result of these orbital configurations.*

5. Canada should expand its support of Space Domain Awareness and open satellite and debris information sharing.

*Commentary: Satellite locations at any given time are of interest to a wide variety of space users, including the astronomical community. Moreover, the brightness of satellites in different bandpasses, their time-dependent brightness variations for a given night, their time above the horizon, and their probability of being in a given telescope observing field are all of relevance to conducting astronomical observations. Characterization of flaring and glaring behaviour by certain satellites might be of particular importance due to the sensitive detectors often used at large-aperture astronomical facilities.*

*Overall, greater investment in Space Domain Awareness (SDA) is required, including fostering collaboration between academic researchers and existing Canadian Space Situational Awareness programs carried out by the Department of National Defence (DND), Defence Research and Development Canada (DRDC), and the Inter-Agency Space Debris Coordination Committee (IADC). Moreover, ensuring open access to orbital information will enhance SDA, with potential benefits to all actors, including the DND. While it is understood that some space assets might be excluded from public catalogues for national security interests, it should be recognized that satellite assets are in principle fully observable to even modest telescopes. Even industry has private entities that monitor satellites via global networks[7]. For this reason, Canada should*

---

[7] Examples can easily be found through quick internet searches.

> *seriously consider supporting open sky initiatives and the open distribution of debris and satellite orbital information[8]. Moreover, supporting open data sharing will enable new opportunities between the DND, industry, and academia in SDA without bias toward building observational infrastructure only at secure DND facilities.*

6. Canada should increase investment in radio astronomy hardware development.

> *Commentary: Canada has a well-established reputation for developing innovative radio astronomy techniques and equipment[9]. However, the increasing amount of satellite-borne interference and the possibility of satellite signals that can damage or destroy receiver electronics through their signal strengths[10] will require new approaches and research efforts for Canada to remain a leader in this research area. The National Research Council of Canada and Canadian universities employ many radio engineers who build sought-after equipment for radio telescopes facilities within Canada and worldwide. Two-world class facilities with significant Canadian involvement are ALMA and CHIME, the latter being led by Canadian researchers and built on Canadian soil[11]. CHORD, a Canadian-led and -sited experiment that builds on the success of CHIME, has recently been allocated C$23M by the Canada Foundation for Innovation. The Canadian radio astronomy community is also seeking substantial involvement with the SKA1 facility, a next-generation international radio observatory. The Canadian Astronomy LRP2020 panel report has specifically identified ALMA, CHORD, and the SKA1 as priorities for next-generation research resources. These radio facilities all rely on sensitive receiver electronics, technologies that Canada has played a major role in developing and using in the construction of radio astronomy observatories.*

7. Canada should increase investment in astronomical data processing initiatives.

> *Commentary: Investment in digital research infrastructure in Canada, including the computational processing power needed to mitigate interference in astronomical observations, has lagged over the past decade. As satellite interference grows, more computationally intensive data reduction strategies will be required for standard astronomical data processing. Machine Learning processes will be critical for removing satellite trails from images, which can be a non-linear process. Such techniques will need concerted efforts in their development. LRP2020 recommended that (1) federal commitments to enhance Canada's digital research infrastructure be maintained; (2) the New Digital Research Infrastructure Organization (NRDIO) support the computational needs for astrophysics; and, (3) the National Research Council's Canadian Astronomy Data Centre (NRC's CADC) continue to be supported. The need*

---

[8] An example of such an effort led by Prof. Moriba Jah is the AstriaGraph database and visualization platform: http://astria.tacc.utexas.edu/AstriaGraph/
[9] https://casca.ca/wp-content/uploads/2020/12/LRP2020_December2020.pdf
[10] See chapter 7 of the Dark and Quiet Skies report, as well as recommendation R9 in Chapter 6.
[11] https://chime-experiment.ca/en

| |
|---|
| *for support and infrastructure will become more severe as interference increases due to mega-constellations.* |
| 8. Canada should with all due speed reassess national laws concerning space liability and, to the extent that is feasible, lead efforts to modernize or bring clarity to the international liability regime. |
| *Commentary: The current international liability regime as applied to space has its foundations in the 1967 Outer Space Treaty and the 1972 Liability Convention[12]. The resulting liability regime gives launching states absolute liability (does not require fault) for damages to property or persons on the ground or aircraft due a space object, whether owned by a state or one of its private entities. Article III of the Liability Convention requires fault for on-orbit damage, in contrast to damage on Earth from space objects. The international liability regime requires that a state seek compensation from another state; there is no direct way for a private entity in one state to seek damages from a private entity in another state through the space liability regime. Moreover, the only time that a claim has been filed under the 1972 Liability Convention is when Canada sought compensation from the U.S.S.R. in response to the Kosmos 954 incident (discussed below). Despite this limitation, national laws could be enacted that would lower barriers for private entities to seek damages through the existing international liability regime, as the situation permits. National legislation could also require that a private space entity that is registered in the state to pay some fraction of the damages that the state would have to pay under the international liability regime as a result of that private entity's actions. Altogether, these could be used to empower space users, such as astronomical observatories, to seek compensation for damages through the existing liability regime should they occur. An example might include a space station broadcast destroying an observatory receiver.*<br><br>*Such changes to national legislation must be coordinated internationally to avoid fostering uneven regulatory regimes or, worse yet, flag-of-convenience states. International coordination may also be required to develop an internationally agreed upon definition for what constitutes damages in the NewSpace era. Canada is in a unique position to explore such options due to its history.*<br><br>*The only incident to test these conventions thus far was in January of 1978, when Kosmos 954 re-entered Earth's atmosphere and spread radioactive waste from the satellite's nuclear reactor over a 120,000 square kilometre region, stretching from Baker Lake in Nunavut to Great Stave Lake in the Northwest Territories. It also extended into the northern regions of Saskatchewan and Alberta.* |

---

[12] Morozova & Laurenava (2021), Oxford Research Encyclopedia of Planetary Science. At: https://doi.org/10.1093/acrefore/9780190647926.013.63

> *Exercising the 1972 Liability Convention and General Principles of International Law, Canada claimed C$6M in compensation from the Soviet Union, asserting that the U.S.S.R. had absolute liability for the damages and the cleanup.  Canada settled in 1981 for C$3M.  As noted above, this incident is so far the only time that the 1972 Liability Convention has been exercised, although it did establish some limited degree of norms[13].  Despite those norms and the resulting settlement, it also exposed potential issues in the convention, including the basic interpretation of the events and the current state of international relations.  For example, the U.S.S.R. blamed the Kosmos 954 failure and its subsequent re-entry on an on-orbit collision (ibid).  It also suggested (initially) that the satellite burned up completely. They then claimed that the radiation hazards from the re-entered material were minimal. Upon recognizing that hazardous material may have been distributed in Canada, the U.S.S.R. offered to help with the clean-up effort.  Canada declined, reflecting in part Cold War tensions.  Instead, the United States aided Canada in the cleanup, costing the U.S. at least US$2M, for which the U.S. never sought reimbursement.  There has been some further discussion as to whether the incident might not have met the damage threshold for triggering the Liability Convention (ibid). The total cleanup effort cost Canada C$14M.*
>
> *Thus, while there is some precedent, it has not been tested in a modern context and should be readdressed nationally and internationally. Canada should lead such discussions at UN COPUOS or by another process in a multilateral forum.*

# Context

The development of space has provided substantial benefits to the global population.  An incomplete list includes advancements and increased reliability in farming, climate studies, navigation, shipping, weather tracking, search and rescue, internet use, and financial transactions. Such development has further made forms of astronomy possible that would otherwise be inaccessible.  Space-based observatories provide, *inter alia*, access to wavelengths of light that are blocked by Earth's atmosphere and the possibility of long-term continuous observations of targets. The rise of NewSpace has further made low-cost missions possible, and has in part enabled Canadian astronomy missions such as MOST[14] and BRITE[15]. Radio astronomy itself was developed in part through attempts to better understand telecommunication systems, including satellites.

With these benefits in mind, it is also important to recognize that the development of Earth orbit has also often been done with multiple unintended consequences. In 1963, the U.S. military in collaboration with M.I.T. Lincoln Labs deployed 400 million copper needles into orbit in an attempt to create an artificial ionosphere for stable long-range communications. This program, called project West Ford, was controversial, and had strong opposition from the astronomical community, but those concerns were largely dismissed[16,17]. The project was also quickly rendered obsolete with the first telecommunication satellites, and subsequently abandoned. While most needles de-orbited as expected (*ibid*), a population remains due to clumping from failed dispersal on orbit. Those needles have the potential to interfere with space observatories as debris, but also as optical flashes in some cases[18].

---

[13] A.F. Cohen (1984), Yale Journal of International Law, **10**. At: https://digitalcommons.law.yale.edu/yjil/vol10/iss1/7/
[14] https://www.asc-csa.gc.ca/eng/satellites/most/default.asp
[15] https://www.asc-csa.gc.ca/eng/satellites/brite/default.asp

This is just one of many examples of unsustainable practices, driven by the desire to achieve something now without properly accounting for risks, often because those risks are distributed among many actors and are not realized until a later date[19]. As satellites increased in numbers and tasks, radio astronomy saw important portions of the electromagnetic spectrum become unusable. The development of the geostationary region, known as the Clarke Belt, has essentially blocked a continuous portion of the sky for many radio astronomy studies. The Iridium satellite constellation (66 satellites) has led to persistent interference with an important frequency range for studying a particular type of emission from evolved stars. This interference is through out-of-band emission, an unintended consequence[20]. Up until now, visible light astronomy has not been significantly affected by satellite activity; but, as will be discussed, the development of so-called mega-constellations[21] will lead to data loss for this area of science, as well.

## Astronomy: Benefits to Society

Astronomy is one of the oldest sciences, is fundamentally connected to our sense of time, and provides some of the most stringent tests of our knowledge of the physical world. Johannes Kepler's mathematical description of planetary motion was one of the keys to the development of Newtonian mechanics and gravity, which for example, are now relied upon for determining satellite orbits. Astronomical observations highlighted the limitations of Newtonian gravity and have been used as tests of General Relativity. Astronomical measurements have revealed the structure of the universe and are used to explore its evolution.

Astronomy and the closely related field of planetary science endeavour to explore whether we are alone in the universe, how common Earth-like planets might be in the Galaxy, and whether, perhaps, there is life elsewhere in the Solar System.

Serendipity is also fundamental to astronomy -- the discovery of what was not known. The universe is full of transient events, which can tell us about the inner workings of stars, show us how the Universe is evolving, reveal the existence of new types of astrophysical objects, and much more.

---

[16] Greenberg, D.S. (1962), *Science*, 135, pg. 30. DOI: 10.1126/science.135.3497.30
[17] Shapiro, I. (1966), *Science*, **154**, pg.1445-1448. DOI: 10.1126/science.154.3755.1445
[18] Mandeville & Perrin (2005), *Advances in Space Research*, **35**, 7. At: https://doi.org/10.1016/j.asr.2004.12.045
[19] Hardin (1968), *Science*, **162**, pg. 1243-1248. DOI: 10.1126/science.162.3859.1243
[20] Dark and Quiet Skies report, ibid.
[21] The term "mega-constellations" is used here to mean a group of approximately a thousand or more satellites that work together as a single system.

The effects of astronomy extend well beyond fundamental knowledge. Indeed, astronomy is a necessity. Near-Earth Asteroids (NEAs) are the remnants of planet formation, but are also bringers of destruction, as the craters on the Moon and Earth remind us. Planetary defence is the detection, characterization, and if needed, mitigation of Earth impact threats by these NEAs. A fundamental challenge is the detection of an impact risk with enough advanced warning to make public policy decisions and if needed to launch a deflection mission. Observatories all over the world are surveying the sky for NEA discovery and tracking[22,23].

Astronomy further has direct and indirect economic benefits, including technology development, data science applications, technical training, and science outreach. The LRP2020 gives a detailed overview of many of such benefits, including estimated monetary benefits. For example, the estimated return on investments in space astronomy alone is approximately a factor of 2.35 (see LRP2020 report).

## Appreciation of the Night Sky -- A Human Connection

Humanity has a deep connection to the stars that extends throughout our history. Watching the sky, sometimes aided by ancient observatories like Stonehenge and Wurdi Youang[24], has allowed people to anticipate seasonal changes and successfully farm and hunt for millenia. Nearly every culture has used the stars for telling stories and passing down vital information in the form of constellations[25] - professional astronomers and most Western stargazers today use the names imprinted on the night sky by Greek and Roman mythologies.

In ancient times, humans everywhere in the world had access to completely dark skies. In stark contrast, today 80% of North Americans cannot see the Milky Way from where they live because of light pollution[26]. The lack of darkness that many people now experience due to urban light pollution has been linked to many physical and mental health issues[27], both in humans[28] and wildlife[29]. But there are still pockets of darkness[30] where urban-dwellers can escape the light pollution and experience skies nearly as dark as those seen by our ancestors. Unfortunately, light pollution from satellites will be a global phenomenon[31] - there will be nowhere left on Earth to experience skies free from bright satellites in orbit.

---

[22] https://www.ifa.hawaii.edu/research/Pan-STARRS.shtml
[23] https://catalina.lpl.arizona.edu/
[24] Norris et al. (2013), Rock Art Research, **30**, 1. Available at: https://arxiv.org/abs/1210.7000.
[25] https://figuresinthesky.visualcinnamon.com/
[26] https://advances.sciencemag.org/content/2/6/e1600377
[27] https://www.ncbi.nlm.nih.gov/pmc/articles/PMC2627884/
[28] https://time.com/5033099/light-pollution-health/
[29] https://www.unep.org/news-and-stories/story/global-light-pollution-affecting-ecosystems-what-can-we-do
[30] https://www.pc.gc.ca/en/voyage-travel/experiences/ciel-sky
[31] https://iopscience.iop.org/article/10.3847/2041-8213/ab8016

Anyone who has ever spent time in a truly dark place staring up at the stars understands the powerful feeling of connection and insignificance this act inspires.  Our lives, our worries, even our entire planet seem so inconsequential on these scales -- a feeling that has shaped literature, art, and culture around the globe.  Seeing the night sky makes it immediately obvious that we are part of a vast and wondrous universe full of countless stars, each of which could host planets with other beings -- some of whom might even be looking back at us across distances so large that light itself takes thousands or millions of years to bridge the gap.  Language even reflects this wonder -- the term "the heavens" invokes a higher power and an afterlife, and likely every culture in the world has similar levels of traditional reverence for the skies.  Connecting to the sky is part of our humanity, and everyone in the world is in very real danger of losing that.  The night sky is a precious resource, and one that we should not take for granted.

## The Effects of Mega-Constellations on Stargazing

Astronomy is one of the few sciences that attracts a large number of amateur practitioners. Amateur astronomy groups are extremely popular around the world - countless people enjoy looking up at the sky using telescopes, binoculars, or their unaided eyes enough to join and attend these groups. The Canada-wide amateur astronomy society, the Royal Astronomical Society of Canada[32], currently has several thousand active members. These groups also play a vital role in engaging with the broader public and inspiring people to explore technical fields.

With the naked eye, stargazing from a dark-sky location allows you to see about 4,500 stars, while from a typical suburban location, you can see about 400[33]. Simulations show that at 52° N (close to the latitudes of Saskatoon, Edmonton, Calgary, and Vancouver) hundreds of Starlink satellites will be visible for a couple of hours after sunset and before sunrise (comparable to the number of visible stars) and dozens of these will be visible all night during the summer months[34] (see Figure 6).  Once Starlink approaches 12,000 satellites in orbit[35], most people in Canada will see more satellites than stars in the sky, unless there are successful mitigation efforts to reduce satellite brightness.

To their credit, SpaceX and Amazon have voluntarily started participating in discussions with professional astronomers on possible ways to mitigate the effects of thousands of bright satellites on observations.  SpaceX did also try a "darksat" coating on one Starlink satellite, though preliminary measurements by astronomers[36] showed that it was only marginally fainter than others[37].  SpaceX has started outfitting Starlink satellites with sunshades ("visors") to decrease their optical brightness[38]. These have shown some promise for reducing naked-eye visibility, although observations by astronomers[39] have shown that they are not as faint as claimed (i.e., they can still be naked-eye visible). In addition, they are highly variable in brightness, something that is especially hard to mitigate in astronomical observations[40]. Their brightness can further lead to non-linear effects in detectors that can be extremely challenging to remove[41].

---

[32] https://www.rasc.ca/
[33] https://skyandtelescope.org/astronomy-resources/how-many-stars-night-sky-09172014/
[34] https://iopscience.iop.org/article/10.3847/2041-8213/ab8016
[35] Starlink has filed for an additional 30,000 satellites.
[36] https://doi.org/10.1051/0004-6361/202037958
[37] https://www.scientificamerican.com/article/spacexs-dark-satellites-are-still-too-bright-for-astronomers/
[38] https://www.spacex.com/updates/starlink-update-04-28-2020/index.html

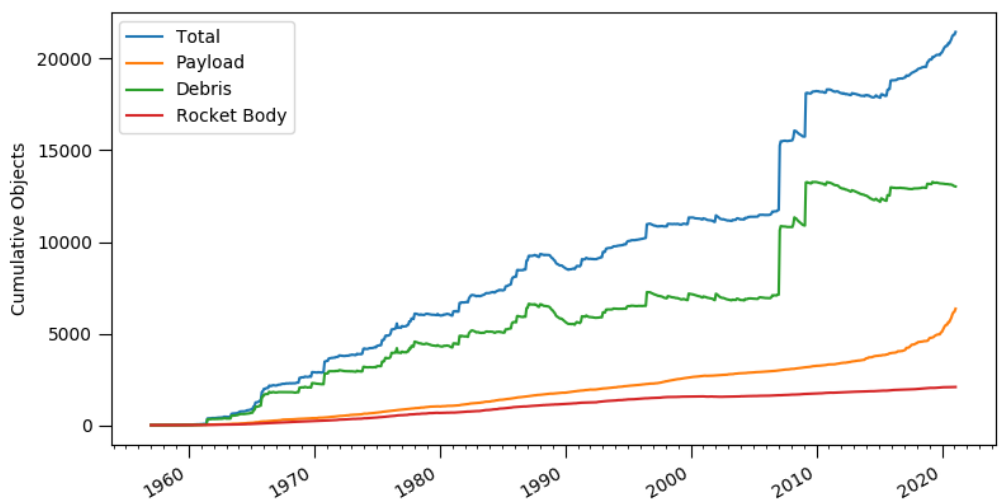

*Fig 1: The cumulative number of objects in orbit over time, including active and defunct satellites (payloads), rocket stages, and debris. The rise in the last couple of years is due to commercial satellites, mainly mega-constellation construction[42]. Data: USSPACECOM.*

## Mega-Constellation Sustainability Concerns

The unobstructed night sky is a finite resource that connects us with traditional ways of knowing and facilitates scientific discovery. Low-Earth Orbit (LEO) is also an extension of the Earth environment, and unsustainable practices in space have ramifications for us on Earth.

Figure 1 shows the growth of the numbers of objects in Earth orbit by type. Up until about 2015, the evolution of the cumulative distribution function (CDF) was driven by fragmentation events. Two major debris events are seen in 2007 and 2009, which are, respectively, the Chinese direct ascent anti-satellite fragmentation test (i.e., a missile launch followed by a kill collision) and the Iridium-33/Kosmos-2251 satellite collision. After that time, growth of the orbital environment has become dominated by NewSpace actors. SpaceX alone has contributed about 1,000 satellites to LEO over the past year -- there are now 5000 catalogued active and defunct satellites

---

[39] https://arxiv.org/pdf/2101.00374.pdf

[40] https://www.nationalacademies.org/event/04-27-2020/decadal-survey-on-astronomy-and-astrophysics-2020-astro2020-light-pollution-rfi-meeting

[41] Tyson et al. (2020), AJ, **160**, id.226

[42] From Boley & Byers, submitted

IWGA Report on Mega-Constellations(payloads) in LEO, up from 3200 about two years ago. SpaceX has approval from the US Federal Communication Commission to place 12,000 satellites into orbit, and has filed for approval for an additional 30,000. OneWeb, Starlink's current closest competitor, has approval for 6,400 satellites at high LEO altitudes of about 1200 km. While such high-altitude shells decrease the brightness of satellites, all other things equal, they also increase the fraction of the sky that solar illuminated satellites can occupy.

To understand better what the growing occupancy of space means for the safe operation of the orbital environment, we show in Figure 2 the resident space object number density as a function of altitude. The plot does not take into account inclination information, so all values are based on averaging over spherical shells. A space resident object's contribution to a given shell is weighted by the time it spends in that shell, using perigee-apogee information from the USSPACECOM[43] catalogue. Construction of the Starlink mega-constellation is already evident at 350 km and 550 km.

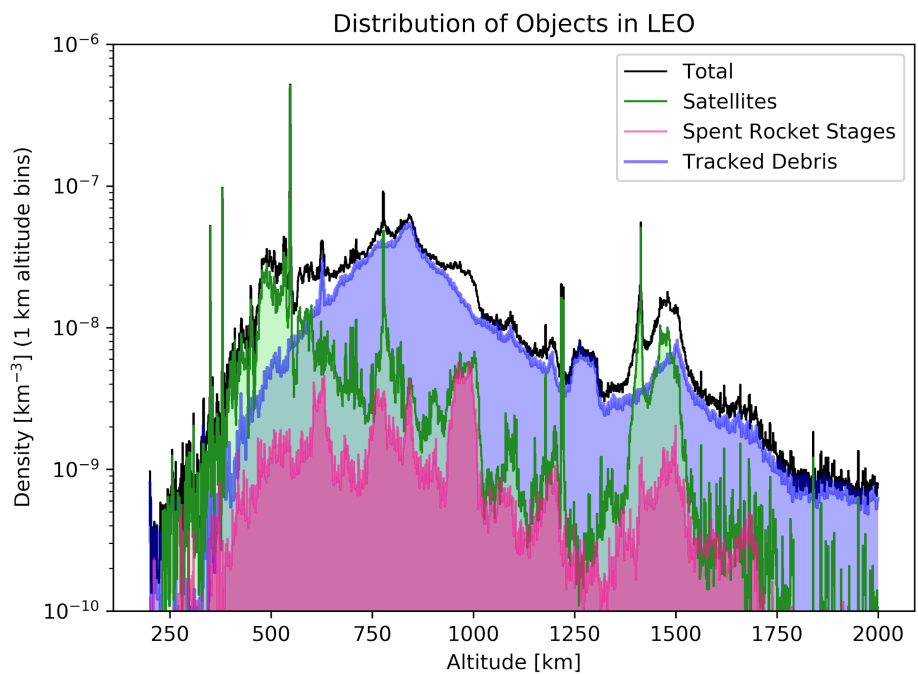

Fig 2: Orbital density profiles for resident space objects in LEO. "Satellites" refers to active and defunct satellites. The tracked debris consists of objects about 10 cm in diameter or larger. When fully constructed, mega-constellations will reach number densities in excess of $10^{-6} km^{-3}$, creating a substantial collision risk with untracked debris, which cannot be avoided[44].

Mega-constellations are taking, at least in part, a consumer electronic model approach to their maintenance. Starlink, for example, expects satellites to be replaced every 5-6 yr. This cadence means that there will be regular disposal of the satellites into the atmosphere, with some satellites having components that will reach the surface. The energy and potential area of such objects are used to evaluate the casualty risk of a satellite re-entry, which is kept below 1:10,000 as a NASA-recommended standard. This risk, however, is applied on a per satellite basis. The original Starlink satellites had approximately a 1:20,000 casualty risk. This means that over a single 5-year cycle, the probability that there will be one or more casualties is 45%. This overall risk was identified (after FCC approval of the original, per satellite risk), and Starlink satellites are now intended to be completely destroyed during atmospheric re-entry, but this highlights how the current single-satellite approach is inadequate for dealing with mega-constellations.

---

[43] Data are available at space-track.org
[44] From Boley and Byers, submitted.

Each Starlink satellite is approximately 260 kg, and made largely of aluminum. If a 5-yr cycle is maintained, then Starlink alone will deposit roughly 2t of aluminum into the atmosphere per day. The meteoroid infall rate is about 54t per day, but only 1% is aluminum. Thus, satellites will exceed the natural deposition of particulates such as alumina in the upper atmosphere[45].

The infall risks are not shared equally by all latitudes. The orbital configuration of the satellites ensures that mid- to high-latitude regions will be areas of orbital caustics, i.e., turn-around points when viewing a satellite's latitudinal variation over an orbit. Satellites become concentrated in those areas, with two prominent caustics over Canadian territory, shown in Figures 3 to 5.

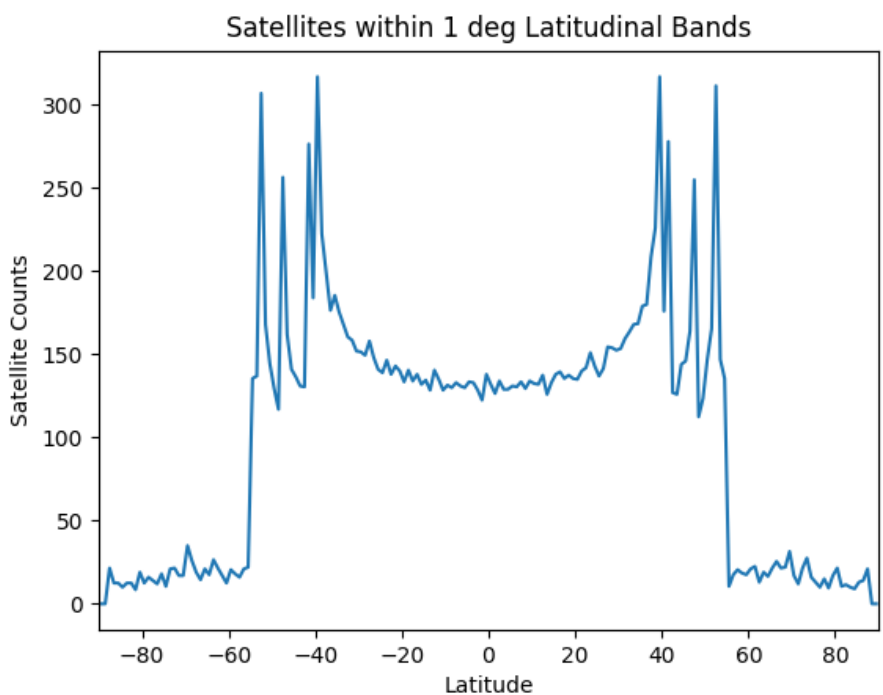

*Fig 3: The number of satellites per 1 degree latitudinal bins, i.e., the number of satellites that are with a 1 degree latitudinal band, extending the band from the surface through LEO. The spikes are the caustics, which are turn-around locations for a satellite's latitudinal variation over one orbit. Canadian territory is crossed by two of these caustics (at approximately 45 degrees and 52 degrees north latitude), which leads to increased satellite infall rates at those latitudes.*

---

[45] Discussed in Boley and Byers, ibid.

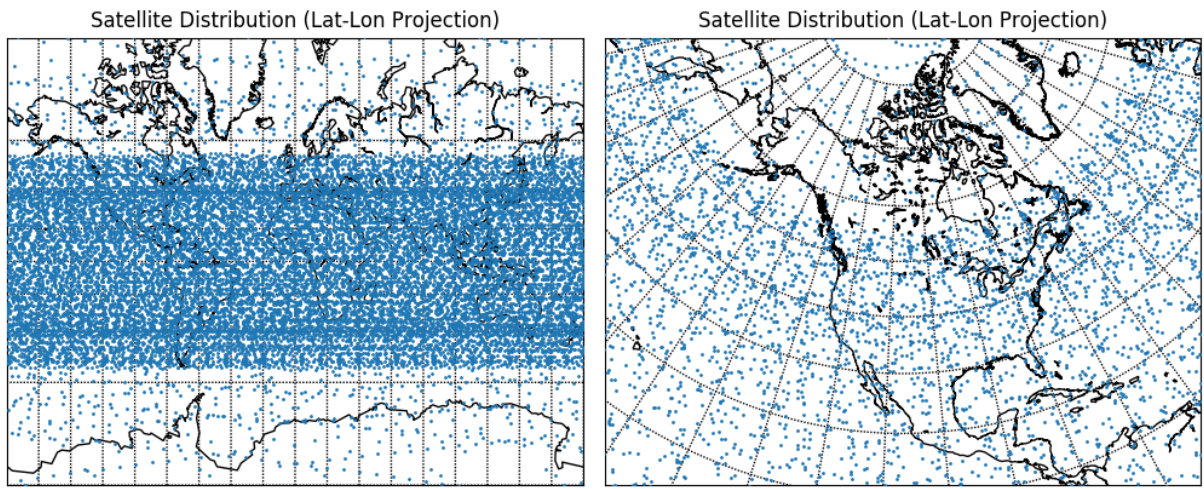

*Fig 4: Distribution of satellites over the Earth, as projected onto latitude and longitude coordinates. The Mercator (left) and Lambert Conformal (right) map projections are both shown.*

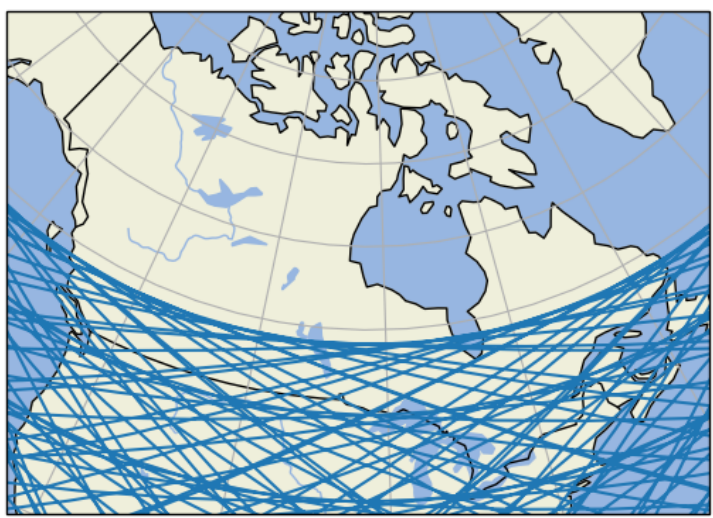

*Fig 5: Similar to Figure 4, but showing select orbital arcs. The caustics are clearly seen as the convergence of multiple orbits at the turnaround points.*

In Figures 6 to 8, we show the impacts of the 12,000 Starlink and the 6,372 OneWeb satellites on the night sky[46]. Each figure shows the total instantaneous number of sunlit satellites above the horizon at a particular time relative to local midnight, as well as the number of satellites that are greater than 30° above the horizon. The number of sunlit satellites depends on the time of year, the latitude of the observer, and the satellite orbital distribution. The satellite's altitude also plays a very important role in the duration of time a satellite remains out of Earth's shadow. The fast variation seen in the figures is a real discreteness effect, i.e., the actual number of satellites at any given moment will vary in addition to an hourly trend. The consequences of the satellite caustics are clearly seen. The black lines in the heat maps correspond to profiles representative of latitudes for the Cerro Tololo Inter-American Observatory (-30º), Maunakea (20º), La Palma (29º), and a latitude generally representing several Canadian observatories (50º). We note that the total number of satellites above the horizon at a given location will always be a potential source of interference for radio telescopes, regardless whether the satellites are illuminated.

---

[46] See also McDowell (2020), *ApJL*, 892, id.L36

It must be recognized that an illuminated satellite in and of itself does not mean that it will be observable to the naked eye, but as discussed above, indications so far suggest that many will be[47,48]. All satellites will be visible to even modest telescopes, and the severity of detector responses (and the overall effects on science data) will depend on the success of brightness and broadcast mitigation efforts.

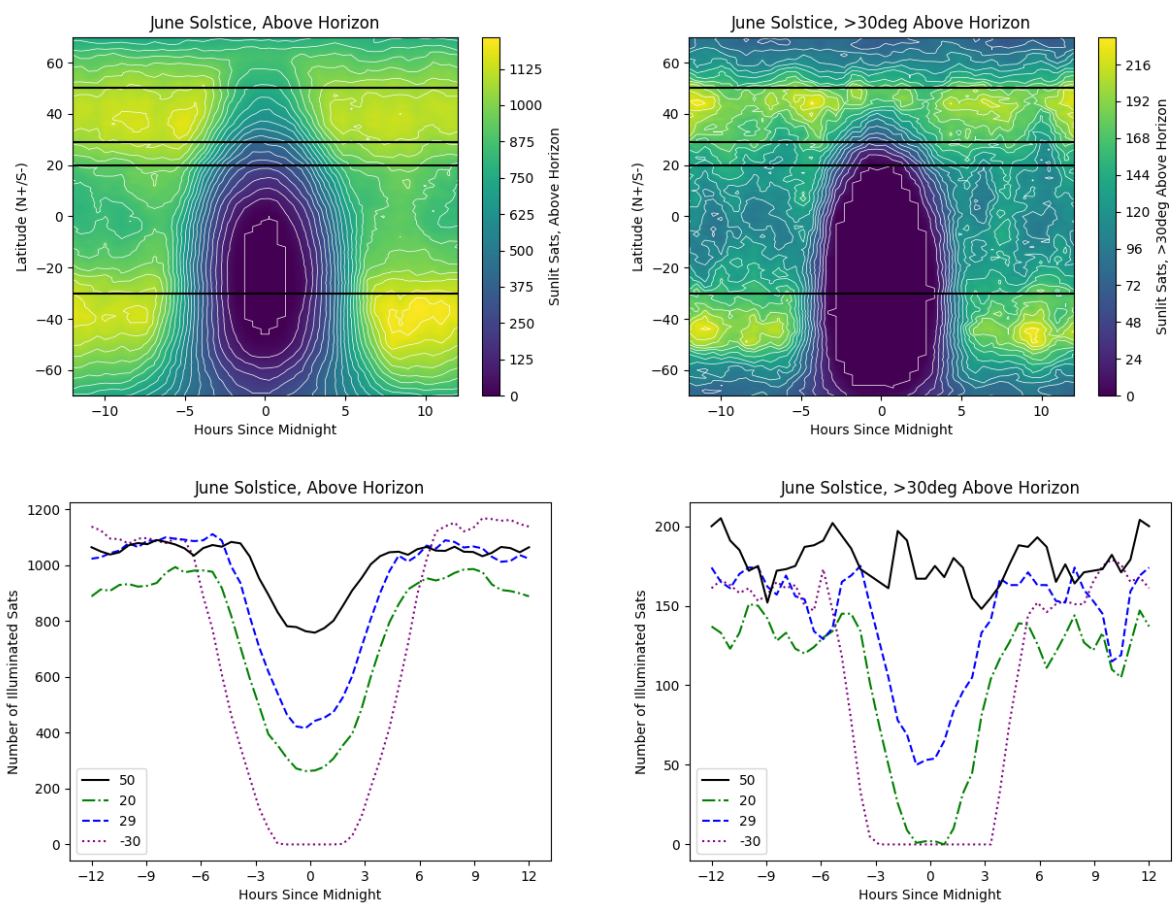

*Fig 6: Number of satellites above the horizon and 30º above the horizon for a range of latitudes during the June Solstice. For Canadian observatories, we can expect there always to be at least 150 illuminated satellites above 30º from the horizon, which is typically regarded as the usable region of the sky for taking high-quality observations (due to the need for light to pass through less atmosphere than closer to the horizon). For radio facilities, there will remain a relatively constant level of satellites above the horizon throughout the day, represented by the daytime sunlit values.*

---

[47] Dark and Quiet Skies Report, ibid
[48] https://arxiv.org/pdf/2101.00374.pdf

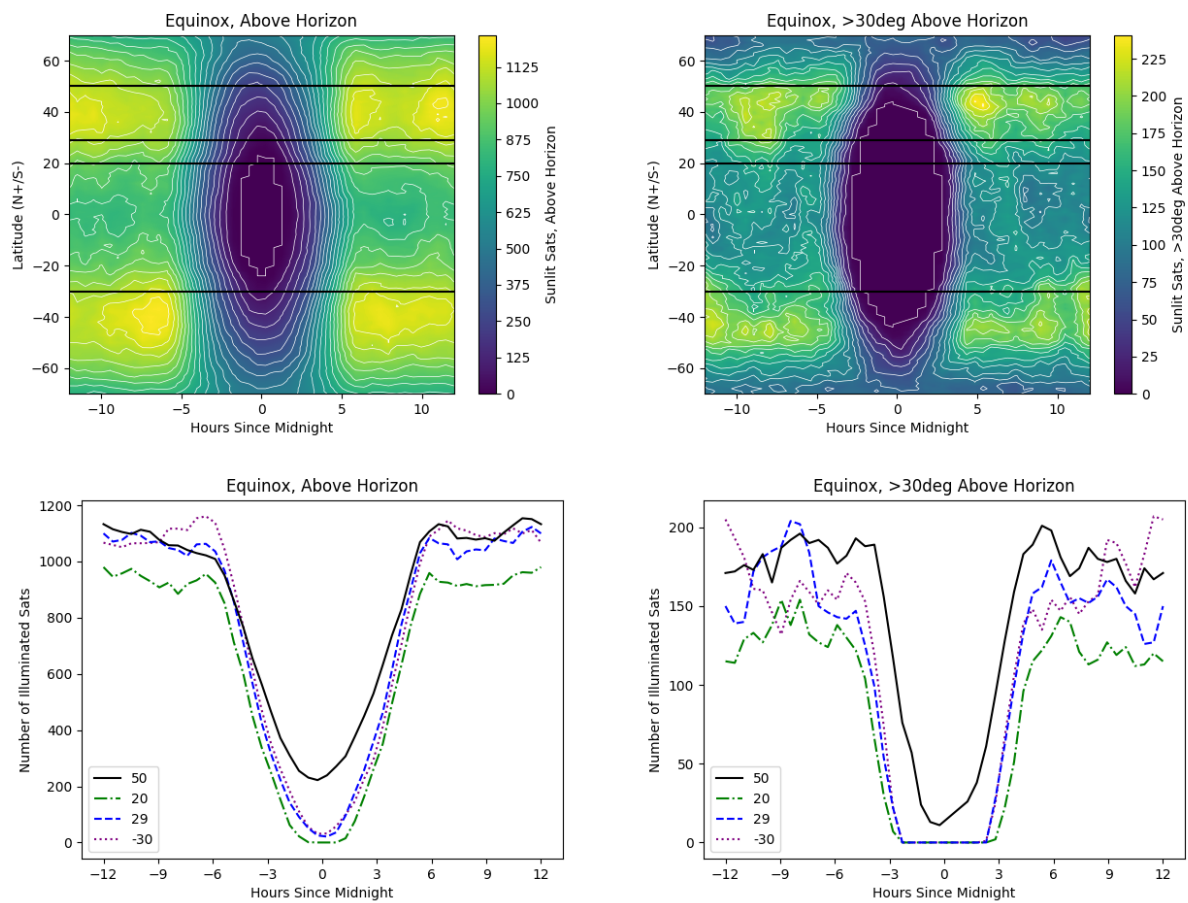

*Fig 7: Similar to Figure 6, but for an equinox.*

Finally, we must emphasize again that mega-constellations with naked eye visible satellites play a role in depleting a shared, finite resource: dark skies. Indigenous traditional knowledge of astronomy is the perfect example of the interrelatedness of elements including natural movements, ebbs and flows, and the interrelationships between the land and sky. Accessing these Indigenous ways of knowing and passing this knowledge to future generations requires access to the unobstructed night sky. In order to follow the UN Declaration on the Rights of Indigenous Peoples[49] and the Truth and Reconciliation Commission Calls to Action, prior to obtaining licensing, launching companies must hold meaningful consultations with Indigenous people and obtain their free, prior, and informed consent.

---

[49] https://www.un.org/development/desa/indigenouspeoples/declaration-on-the-rights-of-indigenous-peoples.html

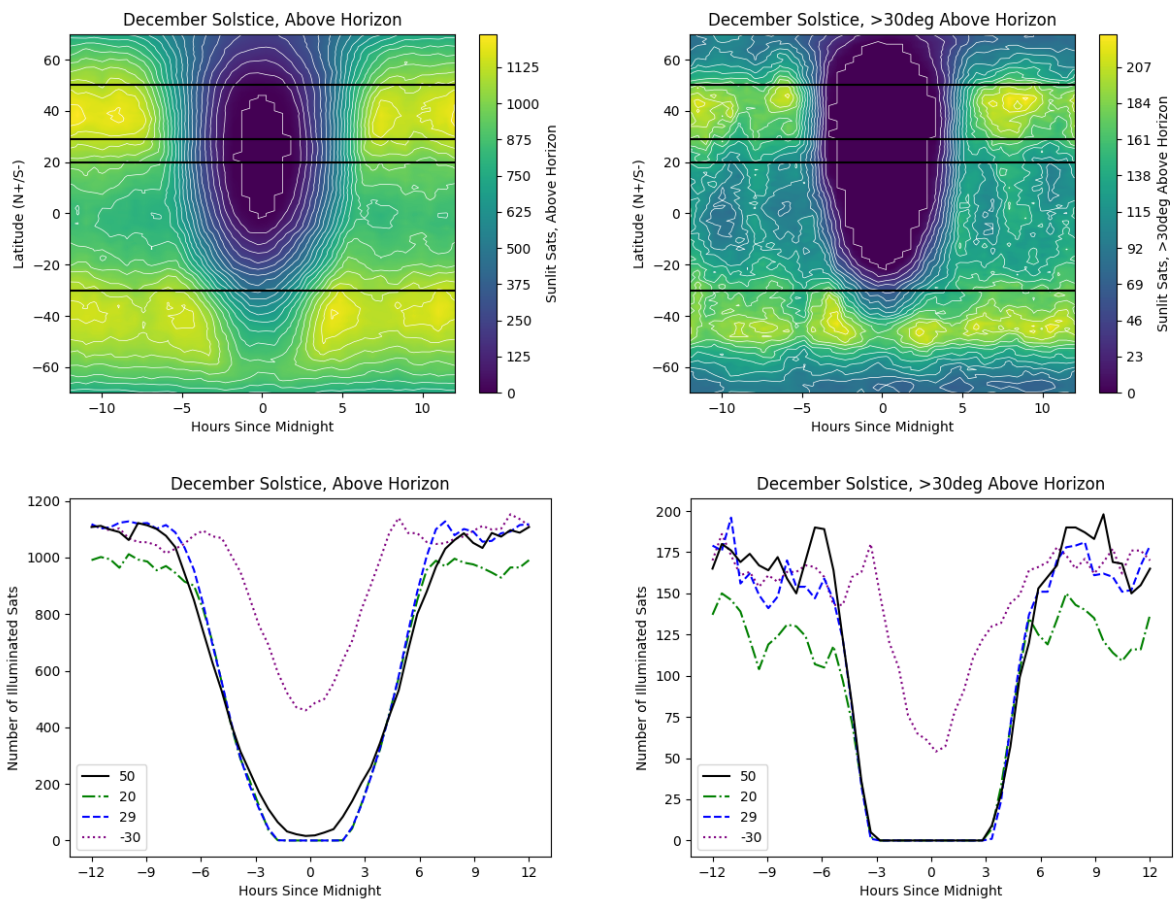

*Fig 8: Similar to Figure 6, but for the December solstice. In this case, the effects of satellites on the night sky are reduced for Canadian latitudes but enhanced for latitudes corresponding to Gemini South (an observatory with Canadian involvement) and the Vera C Rubin Observatory (Canadian involvement a priority).*

## Major Impacts to Canadian Involvement in Optical Surveys

Canada has been a world-leader in wide-field optical astronomy, through projects such as the CFHT Legacy Survey[50] (CFHT-LS) and the Canadian-led Outer Solar System Origins Survey[51] (OSSOS) carried out at the Canada-France-Hawaii Telescope. The CFHT-LS refined astronomers' understanding of the expanding universe while OSSOS discovered hundreds of new icy planetesimals in the solar system. Wide-field imaging covers large areas of the sky and is the most susceptible to interference by the bright trails of satellite mega-constellations[52]. Canadian astronomers hope to build on our legacy of wide-field imaging expertise through involvement in the Legacy Survey of Space and Time (LSST), a revolutionary project which will soon begin to scan the sky every few nights with the Vera C. Rubin Observatory in Chile. LSST participation is a priority listed in LRP2020.  The effects of satellite mega-constellations on LSST observations are still being quantified but are expected to be significant[53], resulting in monetary impacts through scheduling inefficiencies and data loss.  There will also be delays and losses in astronomical discovery and science, as well as in planetary defence through near-Earth asteroid discovery and tracking.

---

[50] https://www.cfht.hawaii.edu/Science/CFHLS/
[51] http://www.ossos-survey.org/
[52] IAU Dark and Quiet Skies report

LRP2020 further recommends that Canada participate in a Very Large Optical Telescope (VLOT) and noted that such participation is the astronomical community's highest ground-based priority. Canada has heavily invested in the Thirty Meter Telescope project, which includes leadership roles in science planning and instrumentation development at National Research Council facilities and Canadian universities, as well as Canadian industry contracts.  The telescope's original site was planned for Maunakea in Hawai'i, although there is potentially an alternate site at La Palma in the Canary Islands[54]. Regardless where the telescope is eventually built, mega-constellations are poised to interfere with observations for any future wide-field (large field of view) instrumentation.  The interference will be more severe at the Canary Islands site due to its higher latitude (see Figures 6-8).

As shown in Figures 6-8, observatories on Canadian soil are also expected to be impacted by mega-constellations, including the Dominion Astrophysical Observatory, the Dominion Radio Astrophysical Observatory, and CHIME in BC; the Rothney Astrophysical Observatory in Alberta; and the Observatoire de Mont-Mégantic in Quebec.

While most analyses are concerned with sunlit satellites, it is necessary to point out that, in addition to potential interference with radio facilities, unlit satellites could also be a problem through occultations, i.e., a satellite passing in front of a source and blocking its light.  Satellite rates on the sky are very fast, so this issue will not affect most observing modes.  Studies that seek to understand very rapid variations in time, however, will be affected.  An example of some concern is the RECON[55] study, which has substantial Canadian involvement.  This international observing collaboration, which includes professional and amateur astronomers, utilizes a network of small-aperture telescopes to search for stellar occultations due to small Solar System bodies.  Since such observations require measuring the brightness of stars many times per second, an ill-timed satellite passage could cause gross inaccuracies in these measurements.  While currently unlikely, the number of satellites on the sky could grow high enough to make these passages occur often.

---

[53] Tyson et al. (2020), "Mitigation of LEO Satellite Brightness and Trail Effects on the Rubin Observatory LSST", AJ, **160**, 5. DOI:10.3847/1538-3881/abba3e

[54] Indigenous rights and consent make the viability of the Maunakea location uncertain. We refer to the LRP2020 report and community papers for a full discussion concerning the history and important associated issues.

[55] http://tnorecon.net/

**This document was prepared jointly by**

**Dr. Aaron Boley**[56]
Canada Research Chair in Planetary Astronomy, University of British Columbia, and Co-Director, Outer Space Institute

**Dr. Samantha Lawler**[57]
Assistant Professor of Astronomy, Campion College at the University of Regina

**With contributions from**

| | |
|---|---|
| Dr. Pauline Barmby | Professor, University of Western Ontario |
| Dr. James Di Francesco | Director, Optical Astronomy, Herzberg Astronomy and Astrophysics Research Centre, National Research Council of Canada |
| Andrew Falle | Research Coordinator and Junior Research Fellow, Outer Space Institute |
| Jennifer Howse | Educational Specialist, Rothney Astrophysical Observatory |
| Dr. JJ Kavelaars | Head, Canadian Astronomy Data Centre, National Research Council of Canada |

**Co-signers**

| | |
|---|---|
| Dr. David Bohlender | DAO Telescope Facilities Manager, Herzberg Astronomy and Astrophysics Research Centre, National Research Council of Canada |
| Julie Bolduc-Duval | Director of Discover the Universe/À la découverte de l'univers, Dunlap Institute for Astronomy and Astrophysics |
| Dr. Dennis Crabtree | Director Emeritus, Dominion Astrophysical Observatory, National Research Council, Herzberg Astronomy & Astrophysics Research Centre |
| Dr. Matt Dobbs | Director, CHORD Observatory; Canada Research Chair in Radio Cosmology and Instrumentation, Department of Physics, McGill University |
| Dr. Sara Ellison | CASCA President, University of Victoria |
| Dr. Bryan Gaensler | Director, Dunlap Institute for Astronomy and Astrophysics, University of Toronto |
| Dr. James E. Hesser | Director (1986-2013), Dominion Astrophysical Observatory and Optical Astronomy, Herzberg Astronomy and Astrophysics Research Centre |
| Dr. Victoria Kaspi | Professor of Physics, Director McGill Space Institute, McGill University |
| Dr. Phil Langill | Director, Rothney Astrophysical Observatory, Department of Physics and Astronomy, University of Calgary |
| Dr. Jason Rowe | Canada Research Chair in Exoplanet Astrophysics, Department of Physics and Astronomy, Bishop's University |

---

[56] aaron.boley@ubc.ca
[57] samantha.lawler@uregina.ca

| | |
|---|---|
| Dr. Luc Simard | Director General, NRC Herzberg Astronomy and Astrophysics Research Centre |
| Dr. Doug Simons | Executive Director, CFHT, Canada-France-Hawaii Telescope |
| Dr. Kristine Spekkens | Canadian SKA Science Director, Royal Military College of Canada |
| Dr. Ingrid Stairs | Professor, The University of British Columbia |
| Dr. Robert Thacker | Former CASCA President, Department of Astronomy & Physics, Saint Mary's University |
| Dr. Kim Venn | Director of the UVic Astronomy Research Center. Director of the NSERC CREATE training program on New Technologies for Canadian Observatories, Canadian member of the Thirty Metre Telescope Board, ACURA Board member, Department of Physics & Astronomy, University of Victoria |

## Appendix: IAU Dark and Quiet Skies Recommendations

The recommendations for national and international policymakers are listed here for convenience. The full IAU Dark and Quiet Skies report is available at this website: https://www.iau.org/static/publications/dqskies-book-29-12-20.pdf

**Recommendations for Science Funding Agencies**

**R31** [For observatory leadership and science funding agencies]
Support provision of funding instruments to help astronomy communities and observatories develop software, hardware and facility mitigations.

**R32** [For science funding agencies, observatories, and astronomical community]
Identify necessary technological developments in telescopes, instruments, detectors, receivers, etc, required to mitigate impacts.

**R33** [For science funding agencies, observatories, astronomy community, and national/regional astronomy societies]
Take steps to evaluate and formalise the impacts on funding instruments (i.e. astronomy grant funding) and capital investments (i.e. telescopes and instruments) and report to political levels of governments.

IWGA Report on Mega-Constellations

**Recommendations on Licensing Requirements**

**R34** [For national space regulators and industry]
Formulate satellite licensing requirements and guidelines that take into account the impact on stakeholders, including astronomical activities, and that coordinate with existing efforts in relation to radio astronomy and space debris mitigation.

**R35** [For the International Telecommunications Union, committees on radio astronomy frequencies, national communications regulators, and national radio frequency managers]
Develop inquiries and recommendations that encourage flexible technology that can better share spectral resources while ensuring protection of sensitive radio astronomy operations. Consider study of and incentives for new transmitter requirements toward a dynamic approach where coordination could be automated and based on the frequency of the scientific observation being taken and the direc-tion in the sky where the radio telescope is pointed. Coupled with dynamic spectrum hopping and other techniques, these types of dynamic models could enhance spectrum efficiency and replace the current static model of quiet zones that assume fixed transmitter requirements based on a given set of parameters. Satellite operators should be encouraged to share the details of their radio systems to a much greater extent than contained currently in public filings with the In-ternational Telecommunications Union or radio spectrum regulators that support their authorization or licensing.

**R36** [For national communications regulators and committees on radio astronomy frequencies]
Formulate licensing requirements that take into account the location of radio quiet zones and radio telescopes, such that satellites can avoid direct illumination of these areas

**Recommendations for National Standards Agencies**

**R37** [For national standards organisations, international standards organisations, and national regulatory agencies]
Develop spacecraft systems and operational standards that take into account the impacts on astronomical science. Areas include reflectivity of surface materials, brightness of space objects, telemetry data, and spurious antenna emissions.

**National economic and space policymakers**

**R38** [For national space regulators, science funding agencies, industry, and space traffic management]
Support the development of space domain decision intelligence collecting data of proposed satellite constellations and existing orbiting space objects, modelling satellites, their operations in the space environment, and estimate uncertainties to assess the impact of satellite constellations on ground-based astronomical observations.

**R39** [For national economic policy makers, national space regulators, and industry]
Investigate policy instruments that account for negative externalities of space industrial activities, including on astronomical activities, and develop incentives and inducements for industry and investors

**Recommendations for Development of International Law**

**R40** [For national space regulators, national space agencies, the astronomy community, industry, national licencing agencies, and COPUOS]
Policymakers are encouraged to contemporaneously develop international agreements, on the one hand, and national laws within their respective legal frameworks, on the other hand, relating to reflected or emitted electromagnetic radiation from satellites, its impacts on science (particularly, but not exclusively, astronomical science), and efforts to mitigate (if not eliminate) the deleterious aspects of such impacts.At both international and national levels, efforts can build upon frameworks in radio astronomy and space debris, informed by this report and by capacity-building and outreach efforts that bring stakeholders together for purposes of discussion and moving policy development forward, such as this international workshop.